\documentclass[prl,twocolumn,showpacs,amsmath,amssymb]{revtex4}
\usepackage{amsfonts}
\usepackage{graphicx}
\usepackage{amsmath}
\usepackage{amssymb}
\usepackage{soul}
\usepackage{color}
 \usepackage[version=3]{mhchem}
\usepackage[colorlinks,bookmarks=false,citecolor=blue,linkcolor=red,urlcolor=blue]{hyperref}

\begin{document}
\title{Magnetic Phase Transition in the Ground-State Phase Diagram of Binary Bose Gases in Optical Lattices}

\author{L. de Forges de Parny$^1$, and Val\'ery Rousseau$^{2}$} 
\email[Corresponding author: ]{laurentdeforgesdeparny@gmail.com} 
\affiliation{$^1$ Universit\'e C\^ote d'Azur, INPHYNI, CNRS, 06000 Nice, France}
\affiliation{$^2$ Physics Department, Loyola University New Orleans, 6363 Saint Charles Ave., New Orleans, Louisiana 70118, USA}

\begin{abstract}
We investigate the ground-state phase diagram of interacting binary Bose gases trapped in two-dimensional optical lattices by means of quantum Monte Carlo simulations. 
Our simulations reveal a magnetic phase transition from a $x-y$ ferromagnetic-order  to a spin insulator
inside the Mott insulating phase with two particles per site for quasi-balanced 
on-site inter- and intra-particle interactions, i.e.,  $U_{\uparrow \downarrow} \lesssim U$.
This 3D-XY transition is characterized by the establishment of a finite local magnetic moment along the $z$-axis,  
ferromagnetic correlations in the $x-y$ plan and by counterflow superfluidity inside the Mott phase.
When decreasing $U_{\uparrow \downarrow}/U$,
this transition merges with the Mott-superfluid transition and becomes first-order.
The merging of the two transitions is investigated with respect to $U_{\uparrow \downarrow}/U$ parameter.
\end{abstract}


\date{\today}

\maketitle

\section{Introduction}

Ultracold atoms loaded in optical lattices are ideal testbed for simulating condensed matter models, such as spin- and (bosonic/fermionic) Hubbard-models \cite{bloch08}.
In these systems, one can reach the limit such that the temperature fluctuations do not play any role  since the associated thermal energy $k_BT$
is much lower than the Hamiltonian energy parameters (i.e., hopping, interactions).
Furthermore, these systems allow a great control of the Hamiltonian parameters by adjusting the intensity of the laser creating the optical lattice  or 
by employing the Feshbach resonances. 
For bosonic atoms, this has allowed the observation of the Mott-superfluid quantum phase transition associated to the spontaneous U(1) symmetry breaking \cite{Greiner02}.

Following this experimental study, many critical phenomena have been studied employing ultracold atoms trapped in optical lattices, to cite a few
the Higgs mode amplitude by simulating an $O(2)$ bosonic model \cite{Endres2012},  
para- to antiferromagnetic phase transition in the Ising model \cite{Simon2011}, 
spin-1 bosonic Mott-Superfluid transitions with $U(1)\times SU(2)$ symmetry \cite{Liu_2016}, 
bosons with competing short- and long-range interactions with $U(1)\times \mathbb{Z}_2$ symmetry  \cite{Landig_2016},
and Bose-Fermi {and Bose-Bose mixtures \cite{Gunter_2006, Ketterle_2009, Ketterle_2020, Inguscio_2008}}.
Hamiltonians with many symmetries offer the possibility to observe many phases and quantum phase transitions in the limit of  zero temperature.
For instance, in the case of spin-1 bosons, theoretical studies have reported two successive transitions when varying the hopping parameter at even filling:
a nematic quantum phase transition inside the Mott insulator phase and the Mott-superfluid transition \cite{Imambekov_2003, batrouni2009, Deforges_2013, Deforges_2014}. 
The nematic  transition spontaneously breaks the $SU(2)$ symmetry whereas the Mott-superfluid transition breaks the $U(1)$ symmetry. 
The relationship between the spin and density degree of freedom is of fundamental importance since it will 
determine the phase diagram and the nature of the quantum phase transitions 
\cite{Altman_2003, Kuklov_Svistunov_2003, Soyler_2009, Capogrosso_Sansone_2010, Kato_2014}.

In this letter, we tackle the question of the {decoupling of the spin and density degree of freedom}
by considering a binary bose gas in a square optical lattice at zero temperature.
{The system under investigation} has the $U(1)\times U(1)$ symmetry that allows two distinct transitions that have 
been extensively theoretically studied   \cite{Altman_2003, Kuklov_Svistunov_2003, Soyler_2009, Capogrosso_Sansone_2010, Kato_2014}.
Using Quantum Monte Carlo (QMC) simulations, 
we unveil the possibility of successively breaking two symmetries, derive the phase diagram 
and elucidate the nature of the transitions employing finite size scaling analysis.
Particularly, we complete the state of the art by revealing a 3D-XY ferromagnetic to spin insulator  
phase transition inside the Mott insulating phase with two particles per site, and 
{study the collapse of the $x-y$ ferromagnetic Mott insulator.}

In the next section, we introduce the model Hamiltonian.  
We discuss our numerical method and the calculated observables in the third section.
Then, we derive the ground-state phase diagram in the fourth section and the 
quantum phase transitions at integer filling in the fifth section.
{The impact of the  decoupling of the spin and density degree of freedom
on the nature of the transition is discussed in the sixth section.}
Finally, conclusions are drawn in the last section.

\section{Binary Bose-Hubbard  Hamiltonian}

{We consider a two-component system of interacting bosonic atoms loaded onto a two-dimensional optical lattice at zero temperature.
Such a binary mixture could be either realized with homonuclear particles in two hyperfine states or with 
heteronuclear species of atoms \cite{Ketterle_2009, Ketterle_2020, Inguscio_2008}. 
The system is governed by the Bose-Hubbard Hamiltonian of two interacting species \cite{Kato_2014}
\begin{eqnarray}
\nonumber
   {\hat{\mathcal H}}  &=&-t\sum_{\langle \bf r,\bf r' \rangle} \left (a^\dagger_{\downarrow, \bf r} a^{\phantom\dagger}_{\downarrow, \bf r'} 
+a^\dagger_{\uparrow, \bf r} a^{\phantom\dagger}_{\uparrow, \bf r'} 
+ {\rm h.c.}  \right ) \\
\nonumber
&&+ \frac{U}{2}  \sum_{\bf r} \left( {\hat n}_{ \downarrow, \bf r} \left ( {\hat n}_{ \downarrow, \bf r}-1\right )  + {\hat n}_{ \uparrow, \bf r} \left ( {\hat n}_{ \uparrow, \bf r}-1\right ) \right)\\
&&+  U_{\uparrow \downarrow}  \sum_{\bf r} {\hat n}_{ \uparrow, \bf r}  {\hat n}_{ \downarrow, \bf r} ~,
\label{Hamiltonian_two_species}
\end{eqnarray}
where operator $a^{\phantom\dagger}_{\sigma, {\bf r}}$ ($a^ \dagger_{\sigma, {\bf r}}$)
annihilates (creates) a boson in the  state $\sigma=\{\uparrow, \downarrow\}$  
on site ${\bf r}$ of a periodic square lattice of size $L\times L$.

The first term in the Hamiltonian is the kinetic term which allows
particles to hop between neighboring sites $\langle {\bf r,r'} \rangle$ with strength $t=1$
to set the energy scale.
The number operator
$\hat{n}_{\sigma,  {\bf r}} \equiv 
 a^\dagger_{\sigma, {\bf r}} a^{\phantom{\dagger}}_{\sigma,
 {\bf r}}$ counts the number of $\sigma$-bosons on site $\bf r$.
$N_\sigma \equiv \sum_{\bf r} \langle  {\hat  n}_{\sigma, {\bf r}} \rangle$ will denote the 
number of $\sigma$-bosons, $\rho_\sigma \equiv N_\sigma/L^2$ the
corresponding density, and $\rho \equiv  \rho_\uparrow +  \rho_\downarrow$ the total density.  
The parameters $U$ and $U_{\uparrow \downarrow}$ are the on-site intra- and inter-particle interactions, respectively.  
These parameters are tunable with the Feshbach resonances \cite{Chin_2010}.
The zero temperature limit  is satisfied if $t, U, U_{\uparrow \downarrow} \gg k_B T = 1/\beta$.
The Hamiltonian, Eq.~\eqref{Hamiltonian_two_species}, has the U(1)$\times$ U(1) symmetry for $U\neq U_{\uparrow \downarrow}$,  
related to the conservation of the atom number in each state \cite{Kuklov_Svistunov_2003}.}

\section{Method and observables}

Hamiltonian, Eq.~\eqref{Hamiltonian_two_species}, is investigated with   
Quantum Monte Carlo (QMC)  simulations  based on  the Stochastic Green Function
algorithm \cite{SGF} with directed updates \cite{directedSGF},
an exact QMC technique that allows for canonical or grand-canonical 
simulations of the system, as well as measurements of
many-particle Green functions.  
In this work, our  simulations in the canonical ensemble are performed  with  balanced populations {$N_{\uparrow} =N_{\downarrow}$}, i.e., 
in the spin sector {$S_{{\rm tot},z}=N_{\uparrow} - N_{\downarrow}=0$}.
Using this algorithm we were able to simulate the system
reliably for clusters going up to $L=16$ with {$N_{\rm tot}=N_{\uparrow} +N_{\downarrow}=512$} particles.
A large enough inverse temperature of  $\beta = 4 L/t$ allows
to eliminate thermal effects {\cite{deforges2017}}.

We calculate the chemical potential, defined as the discrete
difference of the energy
\begin{equation}
\mu(N_{\rm tot})=E(N_{\rm tot}+1)-E(N_{\rm tot}), \label{mu_deltaE}
\end{equation}
 which is valid in the ground-state at zero temperature where the free energy is equal
to the internal energy  $E= \langle  \mathcal {  \hat H}  \rangle $. 
The analysis of the magnetic structure
requires the calculation of the  square of the local magnetic  moment along $z$-axis
\begin{equation}
S_z^2(0) \equiv  \frac{1}{L^2}\sum_{\bf r}\langle {( {\hat n}_{\uparrow, {\bf r}} -{\hat n}_{\downarrow, {\bf r}} )  }^2 \rangle  .
\label{magnetic_local_moment}
\end{equation}
We also calculate the one body Green functions
\begin{equation}
  G_\sigma({\bf R}) =\frac{1}{2L^2}\sum_{\bf r} \langle a^\dagger_{\sigma, {\bf
      r+R}}a^{\phantom{\dagger}}_{\sigma, {\bf r}} + a^\dagger_{\sigma, {\bf
      r}}a^{\phantom{\dagger}}_{\sigma, {\bf
      r+R}}\rangle~,
\label{green_onebody}
\end{equation}
which measure the phase coherence of particles in  state $\sigma$.  
The density of $\sigma$-bosons  with zero momentum  -- here after called the condensate fraction -- is defined by 
\begin{equation}
  C_\sigma =\frac{1}{L^2}\sum_{\bf R} G_\sigma({\bf R})~.
\label{condensate_fraction_QMC}
\end{equation}
The anticorrelated motions of particles, 
defined as the coherent position exchange of particles of different types,
are described by the two-particle counterflow (or anticorrelated) Green functions
\begin{equation}
  G_{a}({\bf R})  = \frac{1}{2L^2}   \sum_{{\bf r}} \langle    
  \hat{a}_{\uparrow, {\bf r+R}}^{\dagger}    
  \hat{a}_{\downarrow, {\bf r+R}}^{\phantom\dagger}    
  \hat{a}_{\downarrow, {\bf r}}^{\dagger}  
  \hat{a}_{\uparrow, {\bf r}}^{\phantom\dagger} + {\rm h.c.}  \rangle. \label{Gapm} 
\end{equation}
If perfect phase coherence is established by means of particle
exchange, $  G_{a}({\bf R})  $ reaches its limiting upper
value equal to  $\rho_\uparrow \rho_{\downarrow} $ at long distances ${\bf R}$.
Due to its definition, $G_{a}({\bf R}) =G_\uparrow({\bf
  R}) G_{\downarrow}({\bf R})$ if there is no correlations between the
movements of particles of up and down spin.
Note that $G_{a}({\bf R})$ measures the ferromagnetic correlations in the $x-y$ plan since 
$  \sum_{{\bf r}} \langle    
  \hat{S}_{{\bf r+R}}^{+}        
  \hat{S}_{ {\bf r}}^{-}  
   + {\rm h.c.}  \rangle
   = 2
     \sum_{{\bf r}} \langle    
  \hat{S}_{x, {\bf r+R}}      
  \hat{S}_{x, {\bf r}}
  + 
    \hat{S}_{y, {\bf r+R}}      
  \hat{S}_{y, {\bf r}}  \rangle$, 
  with 
  $    \hat{S}_{ {\bf r+R}}^{+}   = \hat{a}_{\uparrow, {\bf r+R}}^{\dagger}    
  \hat{a}_{\downarrow, {\bf r+R}}^{\phantom\dagger}   $
  and 
  $    \hat{S}_{ {\bf r}}^{-}   =   \hat{a}_{\downarrow, {\bf r}}^{\dagger}  
  \hat{a}_{\uparrow, {\bf r}}^{\phantom\dagger} $.
The counterflow condensate fraction with zero momentum is given by 
\begin{equation}
  C_a =\frac{1}{L^2}\sum_{\bf R} G_{a}({\bf R}) ~.
\label{anticorr_condensate_fraction_QMC}
\end{equation}
\\

Finally, the superfluid density is calculated with \cite{Ceperley_Pollock}
\begin{equation}
\rho_s= \frac{\langle W^2 \rangle}{4t\beta},
\label{superfluid_density}
\end{equation}
where the winding number $W$ is a topological quantity.
The total superfluid density $\rho_s$ is measured with $W=W_{\uparrow}+W_{\downarrow}$
whereas the counterflow superfluid density $\rho_{sa}$ is measured with $W=W_{\uparrow}-W_{\downarrow}$.

\section{Ground-state phase diagram} 
\begin{figure}[t]
	\includegraphics[width=1 \columnwidth]{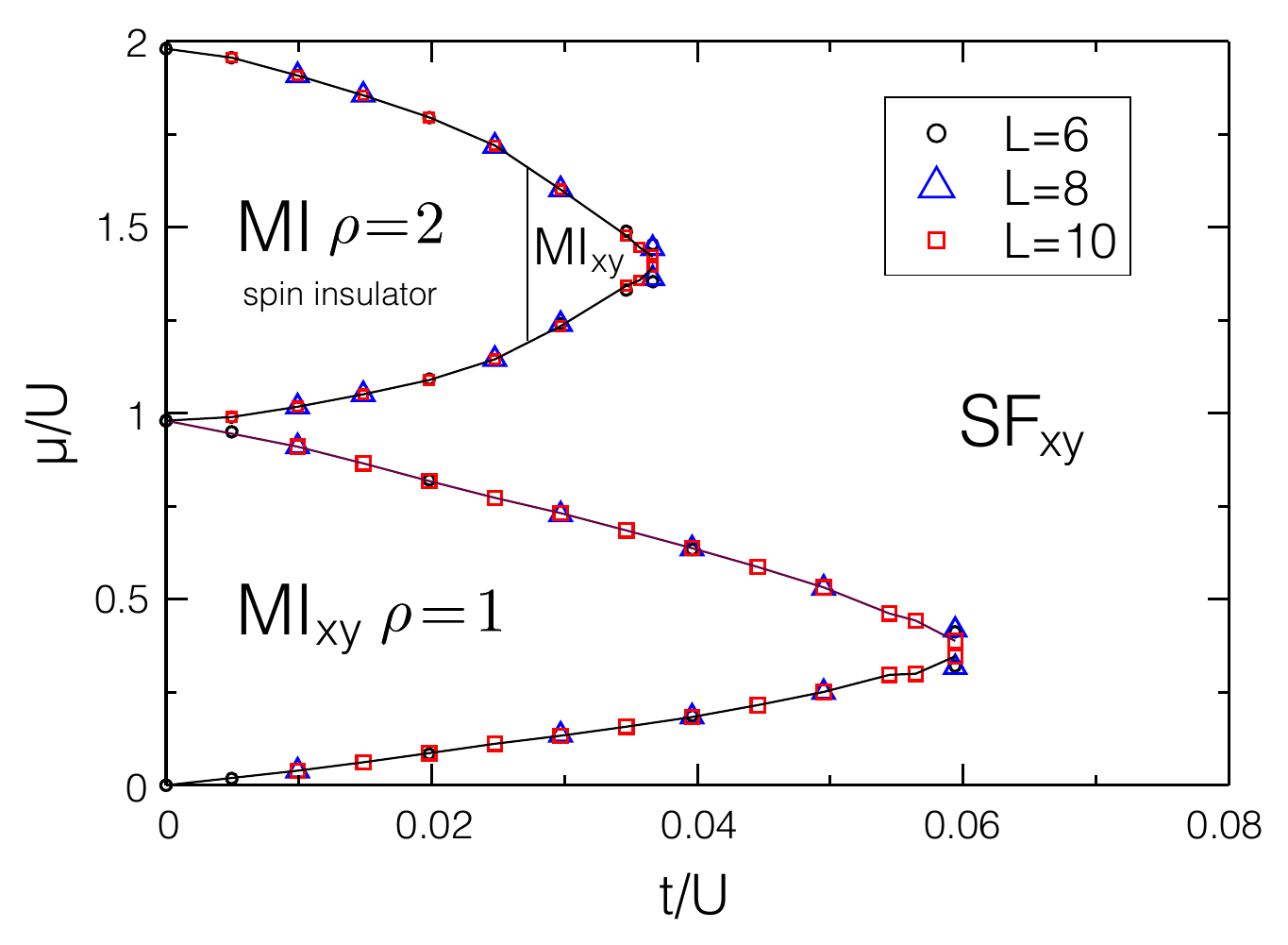}
        \caption {(Color online) {QMC phase diagram of Hamiltonian Eq.~\eqref{Hamiltonian_two_species} with $U_{\uparrow \downarrow}/U=0.98$}, 
        and for system size $L=6$, 8 and 10. A  $x-y$  ferromagnetic order, associated to counterflow superfluidity,  
        is present everywhere except in the spin-Mott insulator phase with two particle per site for $t/U_0<0.028$.
        The spin-Mott insulator   to Mott $x-y$  ferromagnetic (MI$_{xy}$) phase transition belongs to the 3D-XY universality class.
        }\label{Figure1}
\end{figure}

The ground-state phases are characterized by both the mobility of the particle and by their magnetic properties.
In the limit of vanishing hopping {$t/U \to 0$}, the system is in the Mott Insulator (MI) phase with $\rho$ particles per site.
For  $\rho=1$, the basis of  the  Mott  lobe  is {$\Delta \mu = U_{\uparrow \downarrow}$} and completely disappear for {$U_{\uparrow \downarrow}=0$}.
For  $\rho=2$, the basis of  the  Mott  lobe  is {$\Delta \mu = U$} and the  on-site spin-insulating state $|n_\uparrow, n_\downarrow \rangle = |1, 1 \rangle$, 
for which $S_z^2(0) =0$, minimizes the energy such that {$ \langle 1, 1 | \hat {\mathcal { H}}(t=0) |1, 1 \rangle = U_{\uparrow \downarrow}$}.
When turning on the hopping parameter, the Mott phases progressively disappear for the benefit of the superfluid (SF) phase. 
The transition between MI and SF phases is determined by the closure of the density gap, for different system sizes for taking into account the finite size effects, see Figure~\ref{Figure1}.
The density gap $\Delta \mu=\mu_+-\mu_-$ for a system with $N_{tot}$ particles and  energy $E(N_{tot})$ is 
defined by the energy difference in adding, $\mu_+ = E(N_{tot}+1)-E(N_{tot})$, or removing, $\mu_- = E(N_{tot})- E(N_{tot}-1)$, a particle.

According to the magnetic properties, we observe a $x-y$ ferromagnetic order in the SF$_{xy}$ phase, as well as in the  MI$_{xy}$  with $\rho=1$, 
in agreement with Ref.~\cite{Capogrosso_Sansone_2010, Soyler_2009, Altman_2003}.
As we will show, the $x-y$ ferromagnetic-order is also observed for $\rho=2$.
The signature of the $x-y$ ferromagnetic-order is the development of correlations in the $x-y$ plan, i.e., $  G_{a}({\bf R \to \infty }) \neq 0$ (Eq.~\eqref{Gapm}), 
or equivalently, the appearance of a counterflow condensate fraction $C_a \neq 0$ in the thermodynamic limit (Eq.~\eqref{anticorr_condensate_fraction_QMC}).
Another signature of the $x-y$ ferromagnetic order is the appearance of a counterflow superfluidity with $\rho_{sa} \neq 0$ \cite{Kuklov_Svistunov_2003, Capogrosso_Sansone_2010}.

The novelty of our results  lies in the magnetic properties of the phase diagram for $\rho=2$ for which we observe {a magnetic phase transition 
from a $x-y$ ferromagnetic-order  to a spin insulator without symmetry breaking, 
inside the Mott insulating phase,} thus supplementing the state of the art literature \cite{Kuklov_Svistunov_2003, Capogrosso_Sansone_2010, Altman_2003, Soyler_2009}.
The competition between the  spin-insulating state $|\Psi\rangle  = \otimes_{\bf r}  |1, 1 \rangle_{\bf r} $, minimizing the  ground-state  energy for {$t/U\to 0$}, and the $x-y$ 
ferromagnetic order observed in the SF$_{xy}$ phase for {$t \gg U$} leads to a quantum phase transition inside the Mott phase for {quasi-balanced  
on-site inter- and intra-particle interactions, i.e.  $U_{\uparrow \downarrow} \lesssim U$}, see Figure~\ref{Figure1}.
In the following, we unveil and characterize the spin-Mott insulator to Mott $x-y$  ferromagnetic (MI$_{xy}$) phase transition for many {$U_{\uparrow \downarrow}/U$} values.
Note that a similar (singlet-to-nematic)  phase transition has been investigated for spin-1 bosons with vanishing quadratic Zeeman interaction, 
i.e., for $q=0$ \cite{batrouni2009, Deforges_2013, Deforges_2014, Imambekov_2003, Li_2016, snoek04, Deforges_2018}.

\section{Quantum phase transitions at integer filling} 
\begin{figure}[t]
	\includegraphics[width=1 \columnwidth]{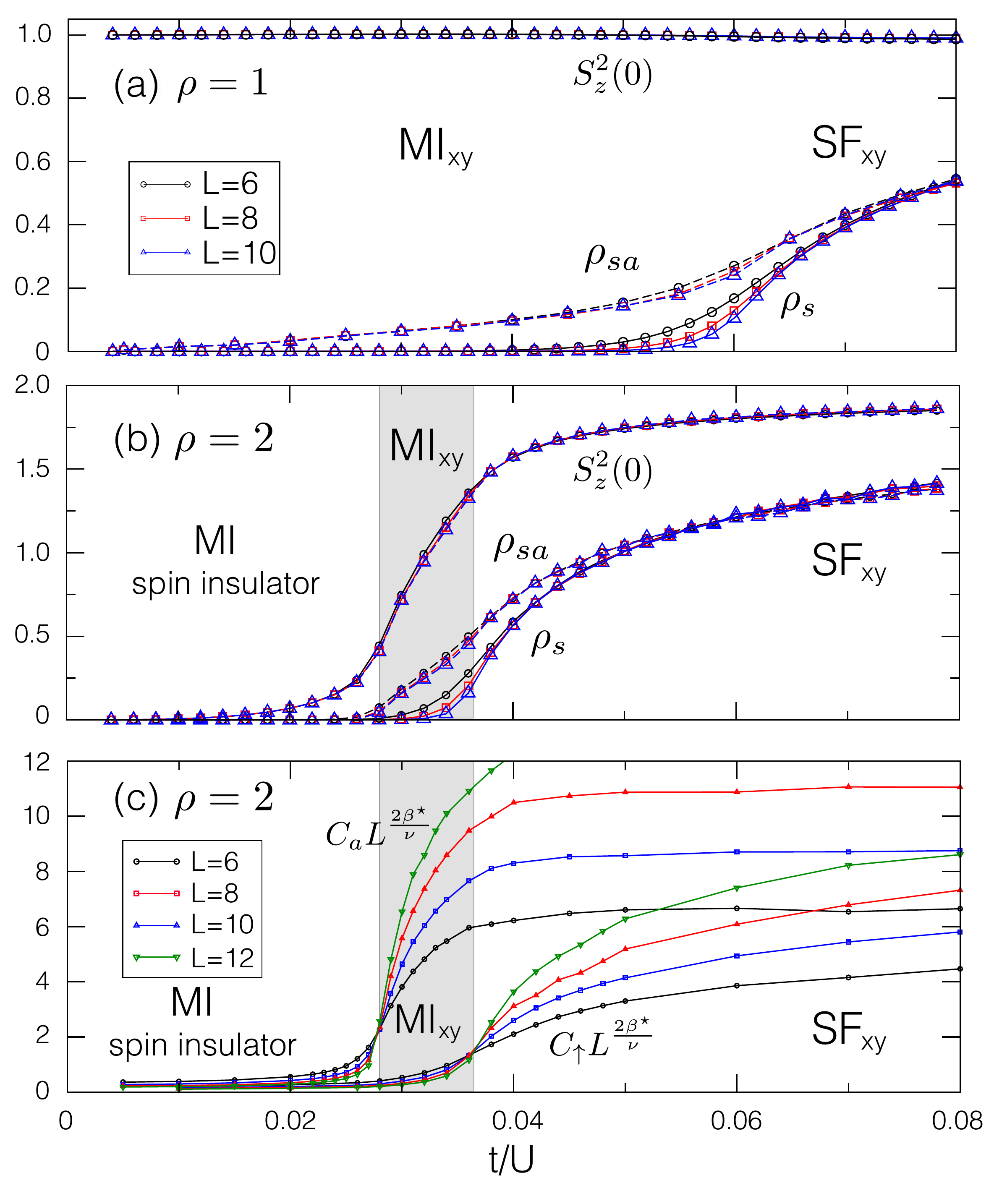}
        \caption {(Color online) {Mott-superfluid transitions at integer filling for  $U_{\uparrow \downarrow}/U = 0.98$ 
        and for system size $L=6$, 8, 10 and 12.} 
        (a) $\rho=1$: the superfluid density $\rho_s$ vanishes in the Mott phase whereas the counterflow superfluid density $\rho_{sa}$ 
        remains finite inside the Mott phase for {$t/U \neq 0$}, and $S_z^2(0) \simeq 1$. 
        These signals indicate a magnetic $x-y$  ferromagnetic order all along the Mott-superfluid  transition.
        (b) $\rho=2$: the  $x-y$  ferromagnetic order, present inside the superfluid phase, disappears in the Mott phase when $\rho_{sa}=0$.
        This phase transition does not match with the Mott-superfluid transition where $\rho_s=0$.
        (c) Finite size scaling of the condensate fractions  $C_{a}$ and $C_\uparrow = C_\downarrow$. 
        We clearly observe two  continuous transitions, located at  {$t_{c1}/U \sim 0.028$ and $t_{c2}/U \sim 0.036$}.
        }\label{Figure2}
\end{figure}

We focus on the quantum phase transitions for one and two particle per site and for a small   {positive $U-U_{\uparrow \downarrow}$} value, i.e., {$U_{\uparrow \downarrow}/U = 0.98$}, see  Figure~\ref{Figure2}.
The signature of the MI-SF transition is a vanishing superfluid density $\rho_s$, around {$t/U \sim 0.06$} for $\rho=1$, 
similarly to the single-specie Bose-Hubbard model \cite{Capogrosso_Sansone_2008}.
For $\rho=1$, we clearly see that the counterflow superfluid density $\rho_{sa}$ remains finite  in both the MI$_{xy}$ and SF$_{xy}$  phases, see Figure~\ref{Figure2} (a).
Also, the local magnetic moment along $z$-axis remains constant $S_z^2(0) \simeq 1$ all along the continuous MI$_{xy}$-SF$_{xy}$ transition.
These signals confirm the presence of a counterflow superfluidity in the MI$_{xy}$ $\rho=1$ phase and the $x-y$ ferromagnetic order  in both the 
MI$_{xy}$ and SF$_{xy}$ phases \cite{Kuklov_Svistunov_2003, Capogrosso_Sansone_2010, Soyler_2009}.
The situation is very different for $\rho=2$, see Figure~\ref{Figure2} (b).
Deep in the Mott lobe, the system adopts the spin insulating state $|\Psi\rangle  = \otimes_{\bf r}  |1, 1 \rangle_{\bf r} $ for which $\rho_{s}=\rho_{sa}=0$ and $S_z^2(0) = 0$.
When increasing {$t/U$} above a critical value {$t_{c1}/U$}, both the counterflow superfluid density $\rho_{sa}$ and  the local magnetic moment along $z$-axis, $S_z^2(0)$, 
become finite for many system sizes, whereas $\rho_s=0$.
This indicates the development of  counterflow superfluidity and $x-y$ ferromagnetic order inside the Mott phase, referred as MI$_{xy}$.
The spin-Mott to MI$_{xy}$ transition can be captured by employing  second-order perturbation theory, for which the Hamiltonian  ${\hat{\mathcal H}}$ maps with \cite{Altman_2003, Kuklov_Svistunov_2003}
\begin{eqnarray}
\label{Effective_Hamiltonian}
   {\hat{\mathcal H}_{eff}}  = - \frac{4t^2}{U_{\uparrow \downarrow}}   \sum_{\langle \bf r,\bf r' \rangle} 
    {\bf {\hat S}}_{\bf r} . {\bf {\hat S}}_{\bf r'}    + (U-U_{\uparrow \downarrow})  \sum_{\bf r}  \left ( S^z_{\bf r}  \right)^2 ~, 
\end{eqnarray}
with 
{${\bf {\hat S}}_{\bf r} =({\hat S^x}_{\bf r} , {\hat S^y}_{\bf r} , {\hat S^z}_{\bf r}  )$} the standard spin-1/2 operator.
The minimization of the free energy leads to a competition between the establishment 
of the spin-insulating state $|\Psi\rangle  = \otimes_{\bf r}  |1, 1 \rangle_{\bf r} $ 
with vanishing fluctuation in the spin densities $\langle  \left ( S^z_{\bf r}  \right)^2 \rangle = 0$  for ${t^2}/{U_{\uparrow \downarrow}} \to 0$ 
and a ferromagnetic order such that 
$\sum_{\langle \bf r,\bf r' \rangle}   \langle  {\bf {\hat S}}_{\bf r} . {\bf {\hat S}}_{\bf r'} \rangle \neq 0$ when increasing ${t^2}/{U_{\uparrow \downarrow}}$.
\begin{figure}[t]
	\includegraphics[width=1. \columnwidth]{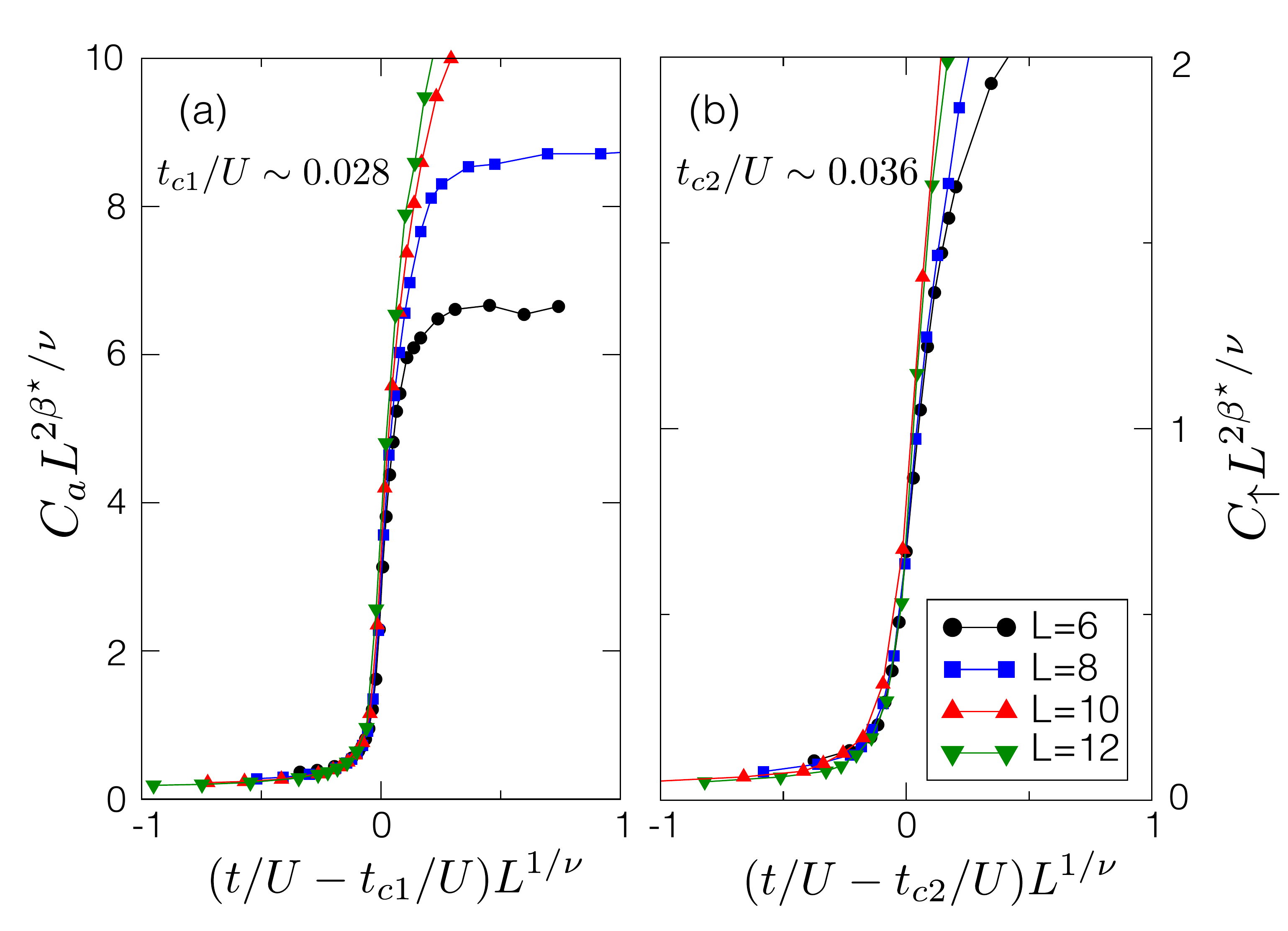}
        \caption {(Color online) Finite size scaling of (a) the counterflow condensate fraction $C_a$ and (b) the condensate fraction $C_\uparrow = C_\downarrow$, 
        {for $U_{\uparrow \downarrow}/U = 0.98$} and $\rho=2$. 
        The two transitions, located at  {$t_{c1}/U \sim 0.028$ and $t_{c2}/U \sim 0.036$},  belong to the 3D-XY universality class
        with critical exponents $\beta^\star=0.3485(2)$ and $\nu=0.67155(27)$ \cite{Campostrini_2001}.       
        }\label{Figure3}
\end{figure}
Our QMC results confirm the existence of the continuous  spin-Mott to MI$_{xy}$ transition, discussed
in Ref.~\cite{Altman_2003}. 
Coming back to Figure~\ref{Figure2} (b), the superfluid phase SF$_{xy}$  with $x-y$ ferromagnetic order appears for $\rho_s\neq0$, when increasing further  {$t/U$}.
Therefore, we observe two successive continuous phase transitions resulting {from the decoupling of the spin and density degree of freedom} for $\rho=2$.
This conclusion is strengthened by  finite size scaling analysis, see Figure~\ref{Figure2} (c).
In the vicinity of the transition point,  the condensates $C_\sigma$ (Eq.~\eqref{condensate_fraction_QMC}) and $C_a$  (Eq.~\eqref{anticorr_condensate_fraction_QMC})
scale as 
\begin{eqnarray}
C_{\sigma, a}  = \xi^{-\frac{2 \beta^\star}{\nu}} f(\xi/L)=L^{-\frac{2 \beta^\star}{\nu}} g(L^\frac{1}{\nu} (t-t_c)/U)~, 
\label{Scaling_relation_3DXY}
\end{eqnarray}
with $\xi$ the correlation length, $L$ the linear system size, 
$f(x)$ and $g(x)$ are universal scaling functions,
and $\beta^\star$ and $\nu$  the critical exponents for the order parameter and correlation length, respectively.
For a two-dimensional system at zero temperature, we expect the transition at integer filling to belong to the 
3D-XY universality class \cite{Fisher_1989}, with exponent  $\beta^\star=0.3485(2)$ and $\nu=0.67155(27)$ \cite{Campostrini_2001}.   
This scaling law and critical exponents allow us to extract the position of the two transitions, i.e., {$t_{c1}/U \sim 0.028$ and $t_{c2}/U \sim 0.036$} for $\rho=2$, see Figure~\ref{Figure2} (c).
The 3D-XY nature of the transitions is also  validated by the collapse of our QMC data in Figure~\ref{Figure3}.

\begin{figure}[t]
	\includegraphics[width=0.9 \columnwidth]{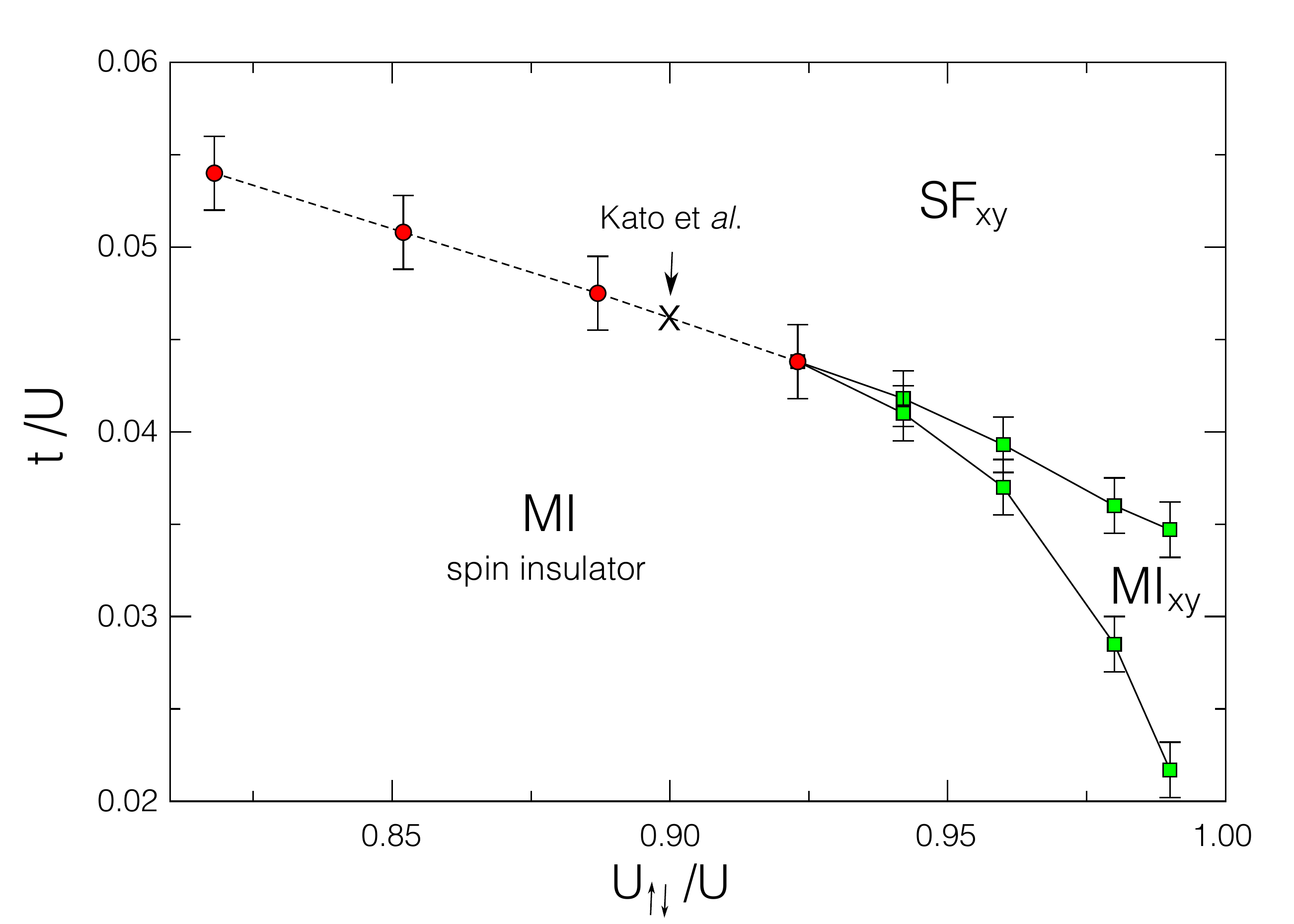}
        \caption {(Color online) {Phase diagram at fixed density $\rho=2$.} 
        {The two successive 3D-XY transitions observed for $0.923 < U_{\uparrow \downarrow}/U < 1$ when increasing $t/U$ (green squares), 
        merge in a single first-order transition (red circles) at the tricritical point for $U_{\uparrow \downarrow}/U\simeq0.923$.
        The first-order transition observed at $U_{\uparrow \downarrow}/U=0.9$ is discussed in detail by Kato et \textit{al.} \cite{Kato_2014}}.
         }\label{Figure4}
\end{figure}
\section{Collapse of the MI$_{xy}$ phase and first-order transition} 
\begin{figure}[t]
	\includegraphics[width=1 \columnwidth]{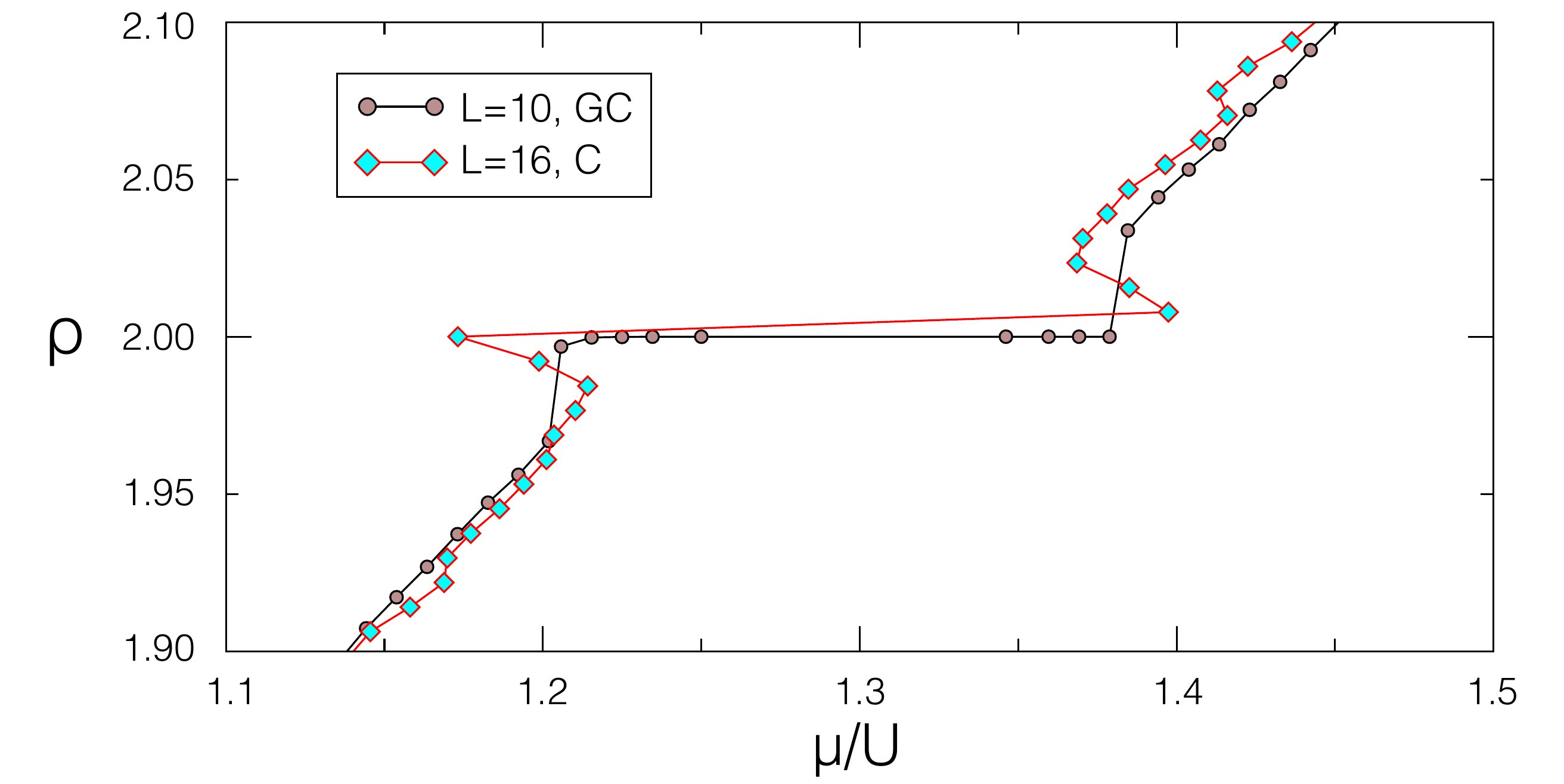}
        \caption {(Color online) Signal of a first-order transition for {$U_{\uparrow \downarrow}/U=0.92$ and $t/U=0.04$} in the 
        canonical (C) and grand-canonical (GC) ensemble.   
        }\label{Figure5}
\end{figure}

The scenario of the two successive phase transitions depicted above for $\rho=2$ dramatically changes when {decreasing $U_{\uparrow \downarrow}/U$}.
For   {small $U_{\uparrow \downarrow} \ll U$}, 
the spin-insulating state $|\Psi\rangle  = \otimes_{\bf r}  |1, 1 \rangle_{\bf r} $ with $S_z^2(0) = 0$
minimizes the ground-state energy in the whole $\rho=2$ Mott phase.
Thus, the MI$_{xy}$ phase vanishes for the benefit of the spin-Mott insulating phase and the {two successive phase transitions} are not observed anymore.
Employing finite size scaling method of Figure~\ref{Figure3}, we find that 
the collapse of the MI$_{xy}$ phase takes place around {$U_{\uparrow \downarrow}/U=0.923$ }
where the two successive 3D-XY transitions merge, see Figure~\ref{Figure4}.
Interestingly, the merge of the two transitions affects the nature of the resulting transition: the spin-Mott insulator-to-SF$_{xy}$ phase transition is first-order for 
{$U_{\uparrow \downarrow}/U < 0.923$ }.
The signal of a negative compressibility $\kappa \propto \partial \rho / \partial \mu$ associated to a first-order transition is found in both 
canonical and grand-canonical ensembles {for $U_{\uparrow \downarrow}/U = 0.92$} \cite{deForges_2011, Deforges_2013}, see  Figure~\ref{Figure5}.
Our results are in agreement with the study of Kato et \textit{al.} \cite{Kato_2014} who finds a first-order transition by means of 
quantum Monte Carlo simulations and effective field theory for {$U_{\uparrow \downarrow}/U=0.9$}. 

\section{Conclusion} 
We have studied the ground-state phase diagram of strongly interacting binary bose gases in square lattices.
Our study is based on QMC simulations, which allow for the calculation of many-body correlation functions, 
and finite size scaling analysis.
In agreement with previous studies, we find a $x-y$  ferromagnetic order, associated to counterflow superfluidity,  
all along the continuous Mott-superfluid transition for one particle per site \cite{Capogrosso_Sansone_2010, Soyler_2009, Altman_2003}.
{The observed scenario for two particle per site is much more interesting since two successive continuous phase transitions 
are observed for quasi-balanced  on-site inter- and intra-particle interactions, i.e., $0.923 < U_{\uparrow \downarrow}/U < 1$}.
{The consequence of the decoupling of the spin and density degree of freedom}
is the appearance of two successive transitions: 
the spin-Mott insulator to  $x-y$  ferromagnetic Mott phase transition and the ferromagnetic Mott-superfluid phase transition.
Our finite size scaling analysis shows that these continuous phase transitions, each breaking the U(1) symmetry,  belong to the 3D-XY universality class.
Experimentally, these transitions can be captured by Ramsey spectroscopy and time of flight measurement \cite{Kuklov_Prokofev_Svistunov_2004, Greiner02}.
Finally, we have shown that the two successive transitions merge into a single first-order transition 
{for $U_{\uparrow \downarrow}/U \lesssim 0.923$}.
In conclusion, the 
{decoupling of the spin and density degree of freedom}
affects the nature of the transition.

\acknowledgments
We thank George G. Batrouni, Fr\'ed\'eric H\'ebert and  Christian Taggiasco for their support 
regarding the numerical facilities (Scylla cluster @INPHYNI). 
We thank Fr\'ed\'eric H\'ebert for his 
critical reading of the manuscript and Tommaso Roscilde for a fruitful private conversation.
We  also thank the anonymous Referee for his relevant comments.
We are also grateful for financial support from the Alexander von Humboldt-Foundation.
Shazbot, Nanu Nanu!

\end{document}